\documentclass[prl,twocolumn,showpacs]{revtex4}

\usepackage{amsmath, amssymb}
\usepackage[dvips]{graphicx}

\begin{document}
\title{Non-Linear Spin Dynamics Of The Electron Liquid In Nanosystems }

\author{ R.N.Gurzhi$^1$,  A.N.Kalinenko$^1$, A.I.Kopeliovich$^1$, P.V.Pyshkin$^1$, S.B.Rutkevich$^2$, A.V.Yanovsky$^1$, A.N.Yashin$^2$}

\affiliation{
 $^1$B.Verkin Institute for Low Temperature Physics and Engineering,
47 Lenin Ave, Kharkov, 61103, Ukraine,
\it \\
{\small \rm
   E-mail:     pavel.pyshkin@gmail.com }\\
   $^2$Joint Institute of Solid State and Semiconductor Physics, P. Brovka Str., 17,
Minsk 220072, Belarus
\it  \\
{\small \rm
   E-mail:     rut@ifttp.bas-net.by}
}

\begin{abstract}We investigate the dynamics of spin-nonequilibrium
electron systems in the hydrodynamic flow regime, when the normal
scattering processes, which conserve the total quasi-momentum of
the system of electrons and quasi-particles that interact with
them, dominate over other scattering processes. We obtain a set of
spin-hydrodynamic equations for the case of an arbitrary electron
spectrum that varies slowly with the coordinates. Solving the
one-dimensional non-linear problem we found the exact solutions
both for the electrical potential in the open-ends circuit and for
the spin-electrical oscillation in an inhomogeneous conducting
ring. As we demonstrate , the oscillation characteristics are
different in the cases of a magnetic ring and a non-magnetic ring
to which the spin polarization was injected. We found  also that a
voltage between the ends of the open circuit may reveal the
presence of an inhomogeneous spin polarization of the electron
density.
\end{abstract}

\pacs{72.25.Hg,
72.25.Mk, 73.40.Sx, 73.61.Ga}

\maketitle
The perspectives of the spintronics, methods of generation and
control of nonequilibrium spin-polarized states in nonmagnetic
conductors are widely discussed recent years, see, e.g., [1]. A
number of promising devices have been proposed. However, usually,
the main attention has been paid to the cases when one may neglect
by the electron-electron scattering.  Some effects of the
electron-electron scattering in the spin dynamics were considered
in Refs. [2-4], but not in the case of the hydrodynamic electron
flow [5] when the electron-electron scattering dominates and the
electron system should be considered as a liquid with its inherent
effects of the stead-state flow and non-steady-state flow.

Meanwhile, a hydrodynamic flow is quite real in nanosystems based
on the semiconductors and semimetal structures as well as in
electron systems over the liquid helium surface [6]. The main
condition is the following. The transport electron mean free path
should be determined by the "normal" collisions that conserve the
total momentum of the system of interacting particles. They may be
electron-electron collisions or electron-phonon collisions (when
phonons are tightly coupled to the electron system and they do not
remove the momentum from the system but they quickly return it
back into the electron system). Note here, that some hydrodynamic
effects were observed experimentally in a high-mobility electron
gas in heterostructures [7].

In the hydrodynamic flow regime, the state of the electron liquid
at the position r is described by two variables: (I) the velocity
$\mathbf{u}(\mathbf{r})$ and (II) the liquid density $\rho
(\mathbf{r})$. When our electron system is spin-polarized, the
densities of the spin components differ from each other and the
electron liquid should be considered as a two-component mixture,
and the density variables have the spin indexes, i.e.,
$\rho_\sigma(\mathbf r)$. Meanwhile, in the leading approximation,
the velocity is the same for all the spin components. The reason
is that frequent collisions between electrons with different spins
form a common drift of the electron system.

Moreover, we have to regard the electron liquid as incompressible
if the geometric size of a conductor is larger than the electron
screening radius which is comparatively small in metals and
heterostructures. Though the total electron density is fixed, the
ratio of the densities of the spin components may vary along the
conductors.

Obviously, the incompressibility and the equal velocities of the
spin components lead to the fact that the total current through
the channel cross section, $I$, does not depend on the coordinate
along the channel and it is distributed between the spin
components in the certain proportion. In Ref. [8] we have found
that in the case of a magnetically inhomogeneous ring, both the
spin polarization and the drift current, $I$, may oscillate
simultaneously with a large decay time. The frequency of the
oscillation is determined by the parameters of the inhomogeneity.

The nature of these oscillations is the following. Due to the spin
inhomogeneity of the ring, the electron drift causes appearance of
a nonequilibrium spin polarization, i.e., an accumulation of the
nonequilibrium densities of the spin-up and spin-down components
occurs (while the total density is conserved). The accumulation
exists until the moment when the drift is stopped due to the
interaction of the nonequilibrium spin density with a field that
induces the inhomogeneity of the electron spectrum. However,
electrons have inertial masses and the process will evolve back.
We call this oscillation process a "spin pendulum." Note, the
existence of the well-known hydrodynamic waves, which frequencies
are less than the plasma frequency, is impossible here due to the
Coulomb interaction. Consequently, "spin-pendulum-like"
oscillations are the only possible oscillations of the system in
this case.

In Ref.[8] we solved a problem of the two-component electron
liquid in a conducting ring in the linear approximation on the
non-equilibrium addition to the electron distribution. However,
when the spin polarization is caused by the spin injection into
the non-magnetic ring (see, e.g. [1]), the linear approximation is
faulty even when the injection level is small: the spin-electrical
oscillation is absent in the linear approximation. In the given
work we obtain the exact solution of the problem both for the
magnetic and non-magnetic conducting channels.

\section{Spin-hydrodynamic equations}

The hydrodynamic equations for the incompressible electron liquid
have been obtained in Ref. [8] by expanding of the kinetic
equation in series in the small hydrodynamic parameters: $l_N/l_i,
l_N/L << 1$, where $l_N$ is the electron mean free path as to the
normal collisions that conserve the total momentum of the system
of interacting particles, $l_i$  denotes the mean free path as to
the all the collisions that dissipate the momentum of the
quasi-particles, $L$ is the size of the conductor.

In Ref. [8] we used some additional approximations: all terms with
the drift velocity $\mathbf u$ have been obtained both in the
linear approximation on $\mathbf u$ and in the main approximation
on the non-equilibrium addition to the electron density. Below, we
generalize the approach for the case of arbitrary values of
deviation on the equilibrium state.

Note, the relaxation processes which lead to decay of the spin
oscillation is of no interest for us here (as we discussed it and
the corresponding decay times in Ref.[8]). Consequently, to obtain
spin-hydrodynamic equations it is enough to write both the law of
conservation of the number of electrons in each of spin states and
the law of conservation of the total quasi-momentum of the
electron system. Thus, we may write
\begin{gather}
\int K d^rp_\sigma = 0, \int \mathbf{p}K d^rp_\sigma = 0 \\
K =\frac{\partial n}{\partial t}+\mathbf{v}\frac{\partial
n}{\partial\mathbf r} - \frac{\partial(\varepsilon +
e\varphi)}{\partial\mathbf{r}}\frac{\partial
n}{\partial\mathbf{p}} \notag\\
n(\zeta)=\left( e^{\zeta/kT}+1\right)^{-1}, \zeta = \varepsilon -
\mathbf{up} - \mu_\sigma \notag
\end{gather}
There are the conditions of solvability of the first approximation
(on the small hydrodynamic parameters) of the quasi-classical
kinetic Boltzmann equation. (The way of derivation of the
hydrodynamic  equation is discussed elsewhere, see, e.g. [9].)  We
assume also that the spin-flip process is vanishing.

Here $\mathbf p$ and $\mathbf v$ are the quasi-momentum and
velocity of an electron, correspondingly, $T$ is the temperature,
$r$ is the dimensionality of the quasi-momentum space and
integration on the $d^rp_\sigma$  is the integration on that
region of the quasi-momentum space which corresponds to the
certain spin state of electrons. $n(\zeta)$ is the main
approximation to the electron distribution function, $\mu_\sigma$
is the spin-dependent chemical potential. The electrical potential
$\varphi$ appears due to the fact that the electron system is in a
nonequilibrium state. Within the quasi-classical approach it is
assumed that the electron energy spectrum depends on the
coordinates and momentum. Here, we have to account a spin index
too:$\varepsilon\equiv\varepsilon_\sigma(\mathbf{p,r})$ (see Ref.
[10]). Using this approach we assume that both the atomic spacing
and the electron wave length are small as to compare with the $L$,
that is the characteristic scale of the electron space-dependent
spectrum.

Here we should note the following. Let the spin splitting of the
electron bands,$\Delta$, is the least characteristic energy. In
the ballistic case, the quasiclassical approach is valid when
$\hbar/\Delta << L/v$.  But, when collisions are frequent, the
"quasiclassical" condition is comparatively mild. Indeed, the
characteristic time of the energy changes is not the $L/v$ as it
was in the ballistic case, but it is the diffusion time $L^2/lv$.

After transforming the integrals in Eq.(1) (see Appendix) we
obtain the following set of equations
\begin{equation}
\frac{\partial\rho_\sigma}{\partial t} +
\mathrm{div}\rho_\sigma\mathbf{u}=0
\end{equation}
\begin{equation}
\frac{\partial g_i}{\partial t} +
\nabla_kP_{ik}+e\rho\nabla_{i}\varphi-\nabla'_iP=0
\end{equation}
\begin{equation}
\rho = \sum_\sigma\rho_\sigma
\end{equation}
Here $\rho_\sigma$ is the spin-dependent current density,
$\mathbf{g}$ is the density of quasimomentum, $P_{ik}$ is the
tensor of the quasimomentum flux (see the corresponding equations
in Ref. 11 for the case of the quadratic spectrum)
\begin{multline}
\rho_\sigma = (2\pi\hbar)^{-r}\int n(\zeta)d^rp_\sigma,\\
\mathbf{g} = (2\pi\hbar)^{-r}\int\mathbf{p}n(\zeta)d^rp,\\ P_{ik}
= g_iu_k+P\delta_{ik}, P=(2\pi\hbar)^{-r}\int
V(\zeta)n(\zeta)d\zeta
\end{multline}
Here $P$ is a hydrodynamic pressure, $V(\zeta)$ is a volume of the
region in p-space that is bounded by the following surface
\begin{equation}
\varepsilon_\sigma(\mathbf{p,r}) - \mathbf{up} - \mu_\sigma =
\zeta
\end{equation}
Note, in Eq.(3)  we wrote the operator $\nabla'$ which means only
that part of the operator $\nabla$ which takes into acount space
dependence of the electron spectrum
$\varepsilon_\sigma(\mathbf{p,r})$, in other words $(\nabla -
\nabla')P = \sum_\sigma\frac{\partial
P}{\partial\mu_\sigma}\nabla\mu_\sigma + \sum_k\frac{\partial
P}{\partial u_k}\nabla u_k$.

As is seen from Eq.(3), the total force, which acts on the
electron system, is caused both by  the electrical field and by
the inhomogeneitiy of the electron spectrum (the last term in
Eq.(3)).

Differentiation of Eq.(6) with respect to coordinate gives (see,
appendix)
\begin{equation}
\frac{\partial P}{\partial\mu_\sigma} = \rho_\sigma,
\frac{\partial P}{\partial\mathbf{u}} = \mathbf{g}
\end{equation}
The first equation corresponds to the well-known thermodynamic
equation (see Ref. 12). Consequently, we may rewrite equation (3)
in the following form
\begin{equation}
\frac{\partial g_i}{\partial t}+ \sum_k(\nabla_kg_iu_k +
g_k\nabla_iu_k) + \sum_\sigma\rho_\sigma\nabla_i\mu_\sigma +
e\rho\nabla_i\varphi=0
\end{equation}

Additionally, we have into account that our electron liquid is
uncompressible and add to Eq. (2) and (8) the following condition
\begin{equation}
\rho(\mathbf{r})=Const(t)
\end{equation}

Thus, we obtain the full set of non-linear equations for functions
$\mu_\sigma(\mathbf r,t)$ and  $\mathbf u(\mathbf r,t)$ which
takes into account that $\rho_\sigma$ and $\mathbf{g}$ are
functions on $\mu_\sigma$, $\mathbf u$ and coordinate $\mathbf r$
(the relation is determined by Eq.(7) where the hydrodynamic
pressure $P$ depends on the $\mu_\sigma$, $\mathbf u$, and,
moreover, it is emplicity dependent on $\mathbf r$, too). Note, in
the linear approximation on deviation of the system on the
equilibrium state (i.e. on $\mathbf u$ and that part of the
$\mu_\sigma$ which depends on the coordinate) this set of equation
is equivalent to the set of the linear equations which has been
obtained in Ref.8.

The set of spin hydrodynamic equations may be simplified
essentially when the electron spectrum is quadratic on $\mathbf
p$. Let $\varepsilon_\sigma(\mathbf{p,r}) = p^2/2m +
\alpha_\sigma(\mathbf r)$. In the sake of simplicity we assume
that the electron mass does not depend on the spin indexes and
coordinates. Thus,  we may write
\begin{multline}
\mathbf{g}=m\mathbf{u}\rho, \rho_\sigma=\rho_\sigma(\mu_{\sigma
c}-\alpha_\sigma), \mu_{\sigma c} = \mu_\sigma + mu^2/2
\end{multline}
In this case we obtain from Eqs. (2),(8), (9) and (10) the
following set of equation
\begin{equation}
\frac{\partial\rho_\sigma}{\partial t} +
\mathrm{div}\rho_\sigma\mathbf{u}=0
\end{equation}
\begin{equation}
\rho\left\{m\left[\frac{\partial\mathbf{u}}{\partial t} +
(\mathbf{u}\nabla)\mathbf{u}\right] + e\nabla\varphi\right\} +
\sum_\sigma\rho_\sigma\nabla\mu_{\sigma c}=0
\end{equation}
\begin{equation}
\rho(\mathbf{r}) = Const(t)
\end{equation}
\begin{equation}
\rho_\sigma = \rho_\sigma(\mu_{\sigma c} - \alpha_\sigma)
\end{equation}
Here we introduce the chemical potential in the co-moving frame of
reference, $\mu_{\sigma c}$. Equation (12) we wrote in the form
which is similar to the Euler equation [11] and used equations
(11) and (12).
\section{Electrical field in the open-end circuit}
In Ref. 8 we solved a one-dimensional spin hydrodynamic problem
for the open-end circuit in the approximation which is quadratic
on the nonequilibrium addition to the chemical potential. Thus, we
found the voltage between the ends of the circuit. Note, it is not
necessary for electron spectrum to be one-dimensional, even though
the whole problem is one-dimensional. The intercoupling between
the voltage and spin polarization provides us a simple way to
reveal the presence of inhomogeneous spin polarization of the
electron density. It is enough to measure the voltage between the
open ends of the circuit. Let us find the relation in the general
non-linear case.

It is clear that current flow is impossible in a circuit with the
open ends, $\mathbf u = 0$, and we have to solve a steady-state
problem. From Eq. (12) we obtain
\begin{equation}
\varphi(L)-\varphi(0) =
-e^{-1}\int\rho^{-1}\sum_\sigma\rho_\sigma(\mu_\sigma,x)\frac{d\mu_\sigma}{dx}dx
\end{equation}
Here we assume that non-equilibrium chemical potentials
$\mu_\sigma(x)$ are the specified functions on the coordinate
along the circuit $x$. $L$ is the length of the circuit.

Let the circuit was magnetically uniform before the spin
injection, i.e. $\rho$  does not depend on the coordinate and
non-equilibrium additions to the spin-dependent electron density,
$\rho_\sigma$, depend on the $x$-coordinate (due to the fact that
corresponding $\mu_\sigma$ are the functions of the $x$). Taking
into account (7) we have
\begin{equation}
\varphi(L)-\varphi(0) = -(e\rho)^{-1}\{P[\mu_\sigma(L)] -
P[\mu_\sigma(0)]\}
\end{equation}

Owning to the incompressibility condition (7), functions
$\mu_\sigma(x)$ in (15) and (16) are not independent, they are
coupled by the following condition:
$\sum_\sigma\rho_\sigma(\mu_\sigma,x)=\rho$. Consequently, we
obtain that $\delta\rho_\sigma=-\delta\rho_{-\sigma}$. As a result
the voltage between the ends (see Eqs. (15) and (16)) is just due
to the spin polarization of the circuit. Specifically, within the
quadratic two-dimensional spectrum ($r=2$) at $T=0$ we obtain
\begin{equation}
\delta P = \frac{\pi (2\pi\hbar)^2}{2m}\left[(\rho_{\uparrow e} -
\rho_{\downarrow e})\delta\rho_\uparrow + \delta\rho^2_\uparrow
\right]
\end{equation}
Here  $\delta P$ is a nonequilibrium addition to the pressure,
$\rho_{\uparrow e},\rho_{\downarrow e}$ are the equilibrium
densities of the spin components ($\rho_\uparrow = \rho_{\uparrow
e} + \delta\rho_\uparrow)$.

To avoid misunderstanding we should note the following. Both the
$x$-dependent non-equilibrium electron densities,
$\delta\rho_\sigma$ , and the spatially-dependent voltage exist
during the limited time after the injection moment. The life-time
scale does not exceed either the spin-flip time or the time of
electron diffusion on the length $L$.

\section{Spin-electrical oscillation in the conducting ring}
Evidently, the electric current may flows in the circuit if we
loop it. As it mentioned above, of special interest is the case
when our conducting channel (ring) is magnetically inhomogeneous,
i.e. when $\rho_\sigma$ depends on the coordinate along the
circuit. (In Ref. 8 we discussed some ways to provide it.)  Let us
demonstrate that corresponding non-linear problem (see
Eqs.(11)-(14)) can be solved exactly.

Really, one can solve the problem due to the following.
Incompressibility of the electron liquid (Eq.(13)) together with
the fact that the number of electrons is conserved (Eq.(11) summed
on $\sigma$ ) lead both to the  homogeneous  distribution of the
current along the circuit, $I =\rho u$, and to the fixed spreading
of the spin components in the total current: $I_\sigma =
\rho_\sigma u = I \rho_\sigma /\rho$ . Thus, we may write the
continuity equation (11) in the form
\begin{equation}
\frac{\partial\rho_\sigma}{\partial t} +
I\frac{\partial(\rho_\sigma/\rho)}{\partial x} = 0
\end{equation}
where $x$ is the coordinate along the ring. The solution of
Equation (18), $\rho_\sigma$, at an arbitrary time-dependent
function $I(t)$ may be written as
\begin{equation}
\rho_\sigma(x,t) =
\rho_\sigma[x_s(x,t),0]\frac{\rho(x)}{\rho(x_s)}
\end{equation}
This equations give us the spatial distribution of the spin
density as a function of the initial distribution
$\rho_\sigma[x_s(x,t),0]$, where $x_s(x,t)$ is the coordinate of
the starting position of an elementary volume of the electron
liquid that is in the position $x$.  The dependence $x_s(x,t)$ is
given by the motion equation
\begin{equation}
\frac{dx}{dt} = \frac{I(t)}{\rho(x)}
\end{equation}

In view of the multiplicative dependence of the movement velocity
on the coordinate and time, it is convenient to introduce a new
variable $y$ which is related to the coordinate $x$ in the
following manner: $dy = \rho(x)dx$. Equations (19) and (20) then
may be written as
\begin{multline}
\rho_\sigma(y,t) = \rho_\sigma(y-Y,0)\frac{\rho(y)}{\rho(y-Y)}, \\
Y(t) = \int_0^t I(t')dt'
\end{multline}
Equation (12) we rewrite in the following form
\begin{equation}
\frac m\rho\frac{\partial I}{\partial t} + \frac 12\frac{\partial
u^2}{\partial x} + \frac
1\rho\sum_\sigma\rho_\sigma\frac{\partial\mu_{\sigma c}}{\partial
x} + e\frac{\partial\varphi}{\partial x} = 0
\end{equation}

Note, the second and the forth terms are the derivatives of the
single-valued functions. Then, after integration over x along the
ring we obtain the following equation for current
\begin{equation}
m\frac{dI}{dt}\oint \frac{dx}{\rho} +
\sum_\sigma\oint\frac{\rho_\sigma}{\rho}\frac{\partial}{\partial
x}\mu_{\sigma c} dx = 0
\end{equation}
(To make our notation clear, below we omit the index "c" on the
chemical potential $\mu$).

Let us demonstrate that equation (23) can be rewritten in the form
of the Newton equation
\begin{equation}
m\ddot Y = -\frac{dU(Y)}{dY}
\end{equation}
The potential energy $U(Y)$ will be found in the general form
below. It is well-known, see, e.g. [13], that the problem of the
one-dimensional motion of a particle may be solved exactly for an
arbitrary potential.

Firstly, taking into account Eq.(21), we rewrite the second term
in Eq.(33) by changing the variables and integrating by parts
\begin{multline}
Q_\sigma = \oint\frac{\rho_\sigma}{\rho}\frac{\partial}{\partial
x}\mu_\sigma dx = -\oint\delta\mu_\sigma\frac{\partial}{\partial
y}\frac{\rho_\sigma}{\rho}dy=
\\ = -\oint\delta\mu_\sigma\frac{\partial}{\partial
y}f_\sigma(y-Y)dy
\end{multline}
\begin{equation}
 f_\sigma(\xi) =
\frac{\rho_\sigma(\xi,0)}{\rho(\xi)}
\end{equation}
Here we have no terms outside the integral, as the integrand is a
periodical function on $y$. $\delta\mu_\sigma$ is the
non-equilibrium addition to the chemical
potential,$\delta\mu_\sigma =\mu_\sigma -\mu_e$, where the
constant $\mu_e$ is the equilibrium part of the chemical
potential.

According Eq.(10), $\rho_\sigma$ are the specified functions of
$\eta_\sigma = \mu_\sigma - \alpha_\sigma$ . Consequently, using
the inverse function $\eta_\sigma(\rho_\sigma)$ we ascertain that
chemical potentials may be written as functions on the $f_\sigma(y
- Y)$ and specified functions on coordinates $\alpha_\sigma$  and
$\rho$
\begin{equation}
\delta\mu_\sigma = \eta_\sigma(f_\sigma(y-Y)\rho(y)) +
\alpha_\sigma(y) - \mu_e
\end{equation}
Taking into account Eq.(27) we may rewrite Eq.(25) for $Q_\sigma$
in the given form
\begin{equation}
Q_\sigma = \frac{\partial}{\partial Y}\oint\int_{a(y)}^{\rho
f_\sigma(y-Y)}[\eta_\sigma(\chi)+\alpha_\sigma-\mu_e]\rho^{-1}d\chi
dy
\end{equation}

Note, integrating Eq.(28) over the variable $\chi$ we have to
assume that $\alpha_\sigma$  and $\rho$  are the task parameters,
and the function $a(y)$, which does not depend on the $Y$, is an
arbitrary function. As our function $Q_\sigma$ in Eq. (28) has a
form of a derivative on the $Y$ we may write the potential $U(Y)$
in Eq.(24) in the form
\begin{multline}
U(Y)=\frac{\displaystyle{\sum_\sigma\oint\int_{a(y)}^{\rho
f_\sigma(y-Y)}\frac{[\eta_\sigma(\chi)+\alpha_\sigma-\mu_e]}{\rho}d\chi
dy}}{\displaystyle{\oint\frac{dx}{\rho}}},\\
 f_\sigma(\xi) =
\frac{\rho_\sigma(\xi,0)}{\rho(\xi)}
\end{multline}
Here we should note that it is enough to posses the
incompressibility condition for the density distribution (see Eq.
(13) ) just in the moment after injection, $\sum_\sigma f_\sigma =
1$.

Within the model of  a "quadratic spectrum", equation (29) is
valid at an arbitrary "shift", $Y$, of the density distribution on
its initial position and for the arbitrary deviation of the
initial distribution $\rho_\sigma(x,0)$ on its equilibrium value,
$\rho_{e\sigma}$,
\begin{equation}
\rho_{e\sigma}(y)=\rho_\sigma(\mu_e-\alpha_\sigma)
\end{equation}

However, we have to take into account that possible deviations are
limited by the case $\rho_\sigma=0$ for one of the spin
components. In that case, variation of the other component leads
to the violation of the incompressibility condition (13) for the
electron liquid. The range of applicability of the theory may be
extended essentially at low level of spin non-equilibrium density
when $\delta\rho_\sigma=\rho_\sigma - \rho_{e\sigma} <<
\rho_{e\sigma}$. In this case the value of the $f_\sigma$ is close
to the equilibrium value, $f_{e\sigma} =
\rho_{e\sigma}(y)/\rho(y)$. Comparing this equilibrium value with
the $f_\sigma(y -Y)$ in Eqs. (26) we conclude that the
non-equilibrium addition is small at an arbitrary $Y$ when both
$f_{e\sigma}$ and $f_\sigma$  are closed to 0,5. In other words,
when spin polarization of the conductor is small both before and
after injection. In the case of the non-magnetic conductors  (or,
if our conductor is very bad magnetic) this condition means a low
level of spin injection: $\rho_\sigma(y,0)/\rho(y) - 1/2 << 1$
(but the amplitude of oscillation may has an arbitrary value). In
the case of the good magnetic, we have an additional condition: $Y
<< L\rho$ , i.e. the shift should be small.

Under the given conditions the set of equations (2), (8), (9) (it
is valid at an arbitrary spectrum) leads to "the Newton equation"
(24) as it was in the case of the quadratic spectrum. Really, in
Eq.(8) we may neglect by the second and third terms (which include
both $\mathbf{g}$ and $\mathbf u$, so they are the second order
terms on the $\mathbf u$) as compared with the first term which is
the first-order term on the $\mathbf u$.

In the main approximation we may write the density of
quasimomentum in the following form
\begin{equation}
g=\mathrm{m}\rho u, \mathrm{m}\rho = (2\pi\hbar)^{-r}\int
p_x^2\left(-\frac{dn}{d\varepsilon}\right)d^rp
\end{equation}

Consequently, in that approximation we may obtain the "Newton
equation" (24) in a manner like the described above, but with the
potential which differs on (29) by the coefficient. However,
$\delta\mu_\sigma$, which is a function of the $\rho_\sigma$ (see
the expression in quadratic brackets in (29), and Eqs. (27),(28)),
have to be expanded it in series on the small deviation of the
density on its equilibrium value
\begin{equation}
\delta\mu_\sigma = \frac{\chi - \rho_{e\sigma}}{\Pi_{e\sigma}}
\end{equation}
Here the equilibrium density of states
$\Pi_{e\sigma}=(\frac{\displaystyle{d\rho_\sigma}}{\displaystyle{d\mu_\sigma}})_e$.
Note, that in the main approximation we may neglect by the
dependence of the density $\rho_\sigma$ on $\mathbf u$ (that
appears in the second order on $\mathbf u$ ) and assume that in
each point of circuit $\rho_\sigma$ is a function on $\mu_\sigma$
only. Integration on $\chi$ yields for the potential
\begin{equation}
U=m\frac{\displaystyle{\sum_\sigma\oint\frac{\rho(y)[f_\sigma(y-Y)-f_{e\sigma}(y)]^2}{\Pi_{e\sigma}(y)}dy}}{\displaystyle{2\oint\frac{\mathrm{m}}{\rho}dx}}
\end{equation}
Integrating (29) on $\chi$ we put $a(y) = \rho_{e\sigma}(y)$.
Generally speaking, the effective mass $\mathrm m$ depends on the
coordinate.

Note, the density of states, $\Pi_\sigma$, does not depend on the
chemical potential for the two-dimensional system with quadratic
electron spectrum at $T=0$. Thus, independent on the level of
deviation of the system on the equilibrium state,   the equations
(32),(33) are the exact expressions, $\Pi_{e\sigma}= 2\pi
m(2\pi\hbar)^{-2}, \mathrm m = m$.

We should stress here, that the oscillation of the spin density is
quite different in the magnetic and non-magnetic conductors, as it
is seen from Eq.(33). In magnetic conductors, when
$\rho_{e\sigma}$ have spin indexes, one may excite the oscillation
without spin injection.

In this case, the oscillation arises even if  $\rho_\sigma(y,0)
=\rho_{e\sigma}(y)$, it is enough to create  the initial current
pulse. The frequency of the small amplitude oscillation, $Y <<
L\rho$ , one may found from Eq. (24), (33)
\begin{multline}
\omega^2 =
{\displaystyle\oint{\left(\frac{\rho_{e\uparrow}}{\rho}\right)'}^2
\frac{1}{\Pi^*_e}dx}\left(\displaystyle\oint\frac{\mathrm{m}}{\rho}dx
\right)^{-1}, \; \\ \frac{\displaystyle 1}{\displaystyle \Pi^*_e}
=\sum\limits_\sigma\frac{\displaystyle 1}{\displaystyle
\Pi_{e\sigma}}.
\end{multline}
Here the "up arrow" corresponds to the one of spin components, a
prime denotes differentiation with respect to $x$. This result is
identical to that obtained in [8].

In the case of non-magnetic conductors, we may rewrite Eq.(33) as
\begin{equation}
U=2m\frac{\displaystyle{
\oint\frac{\rho(y)}{\Pi_e(y)}\left[\frac{\delta\rho_\uparrow(y-Y,0)}{\rho(y-Y)}\right]^2dy}
} {\displaystyle{\oint\frac{\mathrm{m}}{\rho}dx}}
\end{equation}
where $\Pi(\mu)$ is the total density of states. In this case, the
oscillation will be induced by the spin injection. The small
amplitude oscillation will arise if the position $Y = 0$ is closed
to the minimum point of the potential $U(Y)$. The frequency of
oscillation is $\omega  = \sqrt{(d^2U/dY^2)_{min}/m}$ and it is
proportional to the spin injection level.

In summary, we obtain the set of spin-hydrodynamic equations in
the general case of spatially inhomogeneous electron system and an
arbitrary electron spectrum. We found the exact solutions for the
following non-linear one-dimensional problems. Firstly, we found a
voltage between the ends of the open-ended spin polarized circuit.
Within the model of quadratic electron spectrum we reduce the
problem of spin-electrical oscillations in a magnetically
inhomogeneous conducting ring to the problem of non-linear
oscillator. In this case we obtain the expression for the
oscillator's potential energy which is exact at an arbitrary value
of deviation of the system on the equilibrium. In the case when
deviation of the spin density on its equilibrium value is small
(but the amplitude of the oscillation has an arbitrary value) our
results that was obtained for the quadratic spectrum model are
valid for an arbitrary electron spectrum.  The analysis of the
exact results that we obtain allows us to reveal the qualitative
difference between spin-electrical oscillations in magnetic and
non-magnetic conductors.

This work was supported in part by the Joint Ukraine-Byelorussia
Grant No.10.006-F10/51 of Fundamental Researches State Fund of
Ukraine, President of Ukraine Grant for young scientists
GP/F11/0014 and by the NASU Grants No. 1/07-Í, 23/07-Í,  and
3-026/04.
\section{Appendix}
To obtain equation (2) from the equation $\int Kd^rp_\sigma=0$ in
(1), we have perform the following manipulation with the integrals
\begin{multline*}
\int\left[\mathbf{v}\frac{\partial n}{\partial\mathbf{r}} -
\frac{\partial(\varepsilon+e\varphi)}{\partial\mathbf{r}}\frac{\partial
n}{\partial\mathbf{p}}\right]d^rp_\sigma =
\frac{\partial}{\partial\mathbf{r}}\int\mathbf{v}nd^rp_\sigma =\\
=\frac{\partial}{\partial\mathbf{r}}\left[\mathbf{u}\int
nd^rp_\sigma+\int(\mathbf{v-u})nd^rp_\sigma\right]\text{(A1)}
\end{multline*}
Performing the first transformation we integrate the second term
by parts taking into account that
$\frac{\displaystyle{\partial\varepsilon}}{\displaystyle{\partial\mathbf
p}} = \mathbf v$. The last integral in the right-hand part is zero
as it follows from
$$ \int(\mathbf{v-u})n(\zeta)d^rp_\sigma = \int d\zeta
n(\zeta)\oint \frac{dS(\mathbf{v-u})}{|\mathbf{v-u}|}$$ Here $dS$
is the elemental surface area of the surface
$\zeta=\varepsilon-\mathbf{up}-\mu_\sigma$. The integral on $dS$
is zero, $\oint d\mathbf S = 0$, we take into account also that
$\frac{\displaystyle{\partial\zeta}}{\displaystyle{\partial\mathbf
p}} = \mathbf{v-u}$.

Equation (3) follows from the equation  $\int\mathbf pKd^rp = 0$
in (1) by the given transformations
\begin{multline*}
\int p_i\left[\mathbf{v}\frac{\partial n}{\partial\mathbf{r}} -
\frac{\partial(\varepsilon+e\varphi)}{\partial\mathbf{r}}\frac{\partial
n}{\partial\mathbf{p}}\right]d^rp_\sigma = \\ = \sum_k
\frac{\partial}{\partial r_k}\int p_iv_knd^rp_\sigma +
\int\frac{\partial(\varepsilon+e\varphi)}{\partial
r_i}nd^rp_\sigma\\ \text{(A2)}
\end{multline*}

Integrating by parts we assume that near the edges of the
Brillouin zone the occupancy is vanishing and we may neglect by
the integrated term. Let us transform the integral in the
following way
\begin{multline*}
\int p_i(v_k-u_k)nd^rp_\sigma = \int d\zeta n(\zeta)\int dS_kp_i =\\
=\delta_{ik}\int V(\zeta)n(\zeta)d\zeta\qquad\text{(A3)}
\end{multline*}
Here we transform the integration on a closed surface to the
integration on the volume in the p-space that is bounded by the
surface. In other words, the integral is written via the pressure
$P$ in eq. (5). Now, let us find the changes of the volume
$dV_\sigma(\zeta)$ for each of the spin components under the
changes of the chemical potentials, velocity, and coordinates. We
have to differentiate equation (6) under the constant $\zeta$.
Thus, we obtain
\begin{equation*}
|\mathbf{v-u}|dp_\sigma = d\mu_\sigma + \mathbf{p}d\mathbf{u} -
\frac{\partial\varepsilon}{\partial\mathbf{r}}d\mathbf{r}\qquad\text{(A4)}
\end{equation*}
Here $dp_\sigma$ is the displacement normally to the surface of
constant energy $\zeta$. Then, $dV_\sigma(\zeta)=dS_\sigma
dp_\sigma$ and we may write
\begin{multline*}
dP =
(2\pi\hbar)^{-r}\sum_\sigma\int\left[d\mu_\sigma+\mathbf{p}d\mathbf{u}-\frac{\partial\varepsilon}{\partial\mathbf{r}}d\mathbf{r}\right]nd^rp_\sigma
\\ \text{(A5)}
\end{multline*}
Obviously, equation (7) is follows from Equations (A5) and (5),
and taking into account (A2), (A3) we obtain equations (3), (8).
To obtain (3), we have take into account that the third term in
quadratic brackets in (A5) after differentiation of $P$ on the
coordinate  will be cancelled by the corresponding term in the
right-hand part of the (A2).

\end{document}